\documentclass{vgtc}                          

\ifpdf
  \pdfoutput=1\relax                   
  \usepackage{graphicx}                
  \DeclareGraphicsExtensions{.pdf,.png,.jpg,.jpeg} 
\else
  \usepackage{graphicx}                
  \DeclareGraphicsExtensions{.eps}     
\fi%

\usepackage{microtype}                 
\PassOptionsToPackage{warn}{textcomp}  
\usepackage[table,xcdraw]{xcolor}
\usepackage{textcomp}                  
 \usepackage{array,multirow,graphicx}
 \usepackage{float}
\usepackage{mathptmx}    
\usepackage[most]{tcolorbox}
\usepackage{diagbox}
\usepackage[utf8]{inputenc}
\usepackage{times}  
\usepackage{enumitem}
\usepackage{cite}  
\usepackage{multirow}
\usepackage{tabu}                      
\usepackage{booktabs}    
\usepackage{flushend}
\usepackage{hyperref}
\onlineid{0}
\vgtccategory{Research}

\title{Towards Conducting Effective Locomotion Through Hardware Transformation in Head-Mounted-Device - A Review Study\thanks{\textbf{Note}: This paper is a full version of Poster Paper published at IEEE VR 2022, doi: 10.1109/VRW55335.2022.00181. }}

  \author{Y Pawankumar Gururaj\thanks{e-mail: pawankumar.yendigeri@research.iiit.ac.in}\\ 
      \parbox{1.4in}{\scriptsize \centering Center for VLSI and Embedded Systems Technologies\\IIIT Hyderabad, India} %
\and Raghav Mittal\thanks{e-mail: raghav.mittal@research.iiit.ac.in}\\ %
 \parbox{1.4in}{\scriptsize \centering Software Engineering Research Center\\IIIT Hyderabad, India}
 \and Sai Anirudh Karre\thanks{e-mail: saianirudh.karri@research.iiit.ac.in}\\ %
 \parbox{1.4in}{\scriptsize \centering Software Engineering Research Center\\IIIT Hyderabad, India} %
  \and Y. Raghu Reddy\thanks{e-mail: raghu.reddy@iiit.ac.in}\\ %
      \parbox{1.4in}{\scriptsize \centering Software Engineering Research Center\\IIIT Hyderabad, India}
  \and Syed Azeemuddin\thanks{e-mail: syed@iiit.ac.in}\\ %
  \parbox{1.4in}{\scriptsize \centering Center for VLSI and Embedded Systems Technologies\\IIIT Hyderabad, India}     
      }
\abstract{Immersiveness is the main characteristic of Virtual Reality(VR) applications. Precise integration between hardware design and software are necessary for providing a seamless virtual experience. Allowing the user to navigate the VR scene using locomotion techniques is crucial for making such experiences `immersive'. Locomotion in VR acts as a motion tracking unit for the user and simulates their movement in the virtual scene. These movements are commonly rotational, axial or translational based on the Degree-of-Freedom (DOF) of the application.  To support effective locomotion, one of the primary challenges for VR practitioners is to transform their hardware from 3-DOF to 6-DOF or vice versa. We conducted a systematic review on different motion tracking methods employed in the Head-Mounted-Devices (HMD) to understand such hardware transformation. Our review discusses the fundamental aspects of the hardware-based transformation of HMDs to conduct virtual locomotion. Our observations led us to formulate a taxonomy of the tracking methods based on system design, which can eventually be used for the hardware transformation of HMDs. Our study also captures different metrics that VR practitioners use to evaluate the hardware based on the context, performance, and significance of its usage.
} 

\CCScatlist{
  \CCScatTwelve{Hardware}{Communication hardware, interfaces and storage}{Sensor devices and platforms};
  \CCScatTwelve{Human-centered computing}{Human computer interaction (HCI)}{Interaction techniques}
}

\begin{document}
\maketitle
\section{Motivation}
VR products are on the rise. With facebook\textsuperscript{TM} announcing its vision for metaverse\footnote{https://tech.fb.com/connect-2021-our-vision-for-the-metaverse/}, there may be going to be a storm of VR applications into the market like never before. For every VR product, locomotion plays a crucial role in engaging the participant with the VR content. Effective locomotion is achieved if suitable motion tracking methods are best utilized. Unlike external haptic devices, HMDs also offer various avenues to conduct motion tracking to achieve various types of locomotion. However, locomotion capabilities vary between HMDs with a 3-DOF and a 6-DOF. Usually, VR applications that require 3-DOF support can be executed on a 6-DOF supported HMD. However, a VR application that requires 6-DOF support may not be executed on a 3-DOF supported HMD. In contrast, by facilitating some additional hardware on a 3-DOF supported HMD, a closer 6-DOF application can be executed. Such practice is prevalent in the VR practitioner community as most of them upscale or downscale the HMD capacity on supporting motion tracking to manage effective locomotion. Considering these facts, we conducted a literature review on available literature to examine the practices adopted by VR practitioners on how the HMD hardware was transformed from 3-DOF to 6-DOF motion tracking for effective locomotion. 
\\
\newline
\textbf{Why is this study critical? -}  Most VR practitioners who build 3-DOF supported HMD struggle to excel in running rich VR content. However, the slightest hardware transformation will have a significant impact on HMD adoption. This study is scope to understand the hardware transformation of HMDs from 3-DOF to 6-DOF without additional external haptic support.  Learning from our study will assist future targetted HMD developers to develop customizable and configurable HMDs for focused applications. This study paves the way for hassle-free substantial hardware transformation of 3-DOF HMDs on supporting 6-DOF in the future. We also illustrate our observations through a taxonomy of tracking methods through hardware-based HMDs. This taxonomy will also help VR practitioners to plan and transform their HMDs to support additional motion tracking for effective locomotion. 

The rest of the paper is written as follows - Section \ref{2} provides precise details about our review methodology, research questions, search strategy, search string, and its assessment. Section \ref{3} provides our search results and our filtration process. As part of Section \ref{4}, we discuss our observations from the extracted literature, including evaluation methods, metrics, and taxonomy. Section \ref{5} discusses the threats to the validity of our study. Section \ref{6} provides details about related work and along with the conclusion.

\section{Study Step}\label{2}
We conducted our systematic review study by considering the guidelines proposed by Kitchenham et al. \cite{KitchenhamBBTBL09}. As part of our review, we utilized the PICOC (Population, Intervention, Comparison, Outcome, and Context) method to establish our study's context and relevance \cite{experimentation}. This also helped us design our research questions, search string, and search protocol. Table \ref{tab:picoc} illustrates the PICOC details of our study. 

\begin{table}[htbp]
\centering
\begin{tabular}{|c|l|}
\hline
\textbf{Criteria}             & \multicolumn{1}{c|}{\textbf{Description}}                                                                     \\ \hline
\textit{\textbf{Population}}   & \begin{tabular}[c]{@{}l@{}}For VR HMD users willing to switch\\ locomotion from 3-DOF to 6-DOF\end{tabular}     \\ \hline
\textit{\textbf{Intervention}} & \begin{tabular}[c]{@{}l@{}}Motion tracking methods for VR \\ locomotion\end{tabular}                          \\ \hline
\textit{\textbf{Comparison}} &
  \begin{tabular}[c]{@{}l@{}}Comparison between tracking methods \\ based on hardware requirement, working,\\ performance and target application\end{tabular} \\ \hline
\textit{\textbf{Outcome}}      & \begin{tabular}[c]{@{}l@{}}Studies that employed motion tracking \\ methods for locomotion in VR\end{tabular} \\ \hline
\textit{\textbf{Context}}      & \begin{tabular}[c]{@{}l@{}}Academia, VR community and other \\ empirical studies\end{tabular}                 \\ \hline
\end{tabular}
\caption{PICOC details our Review Study}
\label{tab:picoc}
\end{table}

\subsection{Research Questions}
The primary objective of our review study is to summarize the motion tracking methods served by HMDs for conducting effective locomotion in VR applications. Below research questions are expressed to capture the insights of our objective. 
\begin{itemize}
\item \textbf{RQ1}-What types of motion tracking methods are operated for head tracking by an HMD for VR applications?
\item \textbf{RQ2}: What are the hardware components required for transforming a VR HMD from 3-DOF to 6-DOF motion tracking?
\item \textbf{RQ3}: What are the metrics practiced to evaluate the effectiveness of motion tracking methods after transforming the VR HMD from 3-DOF to 6-DOF?
\end{itemize}
\subsection{Search strategy}
We used our research questions to deduce our search strategy. We first created a list of keywords that are relevant to the research questions. We later generalized the keywords by streamlining the scope of the review. We finalized the search string by considering all possible synonyms and have divided them into three parts, i.e., \textit{$S_1$, $S_2$,} and \textit{$S_3$}. Below is our final Search string:
\\
\begin{tcolorbox}[]
\textbf{\textit{$S_1$:}} ``Virtual Reality'' \textbf{OR}   ``VR''
\end{tcolorbox}

\begin{tcolorbox}[]
\textbf{\textit{$S_2$:}} ``Head mounted device''  \textbf{OR} ``Head mounted display'' \textbf{OR} ``HMD'' \textbf{OR}  ``Head-mounted display'' \textbf{OR} ``Head-mounted device'' \textbf{OR} ``headset'' \textbf{OR} ``display'' \textbf{OR} ``projection''
\end{tcolorbox}

\begin{tcolorbox}[]
\textbf{\textit{$S_3$:}} ``Degree of Freedom" \textbf{OR} ``Degrees of Freedom" \textbf{OR} ``DOF" \textbf{OR} ``3 DOF" \textbf{OR} ``3-DOF" \textbf{OR} ``3DOF" \textbf{OR} ``6 DOF" \textbf{OR} ``6-DOF" \textbf{OR} ``6DOF" \textbf{OR} ``motion tracking" \textbf{OR} ``motion-tracking" \textbf{OR} ``head tracking" \textbf{OR} ``head-tracking"
\end{tcolorbox}

Overall our search string is defined \textit{$S_1$} \textbf{AND} \textit{$S_2$} \textbf{AND} \textit{$S_3$}.
\\
\newline
The search string is divided into three parts to carry an organized filtration. The scope of search statement $S_1$ is limited to the abstract of the research paper only. Both $S_2$ and $S_3$ are used to search across the full text of the research paper. We worked with our peer-researchers at our research center over group discussions to address the necessity of each keyword described as part of $S_1$, $S_2$, and $S_3$. We conducted multiple iterations to arrive at a finalized search string. These iterations include a severe review on synonyms, a keyword's relevance, and additional reasoning on search interval to make the search more reasonable. As part of our initial search, we divided our search interval into two periods, i.e., between 2000 - 2010 and between 2011 - 2021. Our cumulative search outputs show that the period between 2000 and 2010 does not provide any significant research contribution. After a three-fold search review by individual peer-researchers, we concluded that the research contribution between the period 2000 and 2010 is either obsolete or not relevant to the current maturity of the VR domain. Thus, we limited our search period between 2011 and 2021 for the essence of a better review. Our review study includes papers published until August 2021.
\subsection{Search Quality Assessment}
We designed a set of ten interrogative questionnaires to aid our review to filter the research papers based on their relevance, reliability, and nature of the study. This questionnaire awards a Yes or No, i.e., 1 or 0 as a score, where yes represents review consideration and No for ignore for review consideration. For a given paper, it requires a score of 6 for review consideration. Our quality questionnaires are explained as follows: 
\begin{itemize}[noitemsep,nolistsep]
    \item Is the tracking methodology novel or follow-up research?
    \item Is there clarity in explaining the objective of the research?
    \item Is it possible to realize the study as a real-life application?
    \item Was the application/necessity of the method addressed in the paper?
    \item Is the data provided in the study addressing objectives of the research?
    \item Is the motion tracking method validated using a study?
    \item Was the validation technique for motion tracking explained appropriately with description and reference?	
    \item Is the information provided enough to replicate the design?
    \item Is the study of value for further research?
    \item Is there mention of findings, limitations, future scope, or discussions in the paper?
\end{itemize}

Apart from the search quality assessment, we employed below inclusion and exclusion criteria to further filter our search output.
\\
\newline
\textbf{Inclusion Criteria} - Only research papers written in English are considered for our study. Only papers published between 2010 and 2021 are considered. The study that provides transparent information about the design, implementation, and evaluation of motion tracking techniques in VR is considered. Papers that discuss and authenticate the motion tracking accuracy based on some user-study are considered. 
\\
\newline
\textbf{Exclusion Criteria}  Studies that are not available in Full Text are not considered for review.  Research contributions published as articles, magazines, review Notes, datasets, archives, books, book chapters, reference works are excluded from the study as they are informal and incomplete in regards to the goal of our search. Studies involving motion tracking using external haptics, external controllers, or objects are excluded from our study.  Paper without proper study to validate the hardware is excluded.

\section{Results} \label{3}
\begin{figure}[htbp]
    \centering
    \includegraphics[scale=0.45]{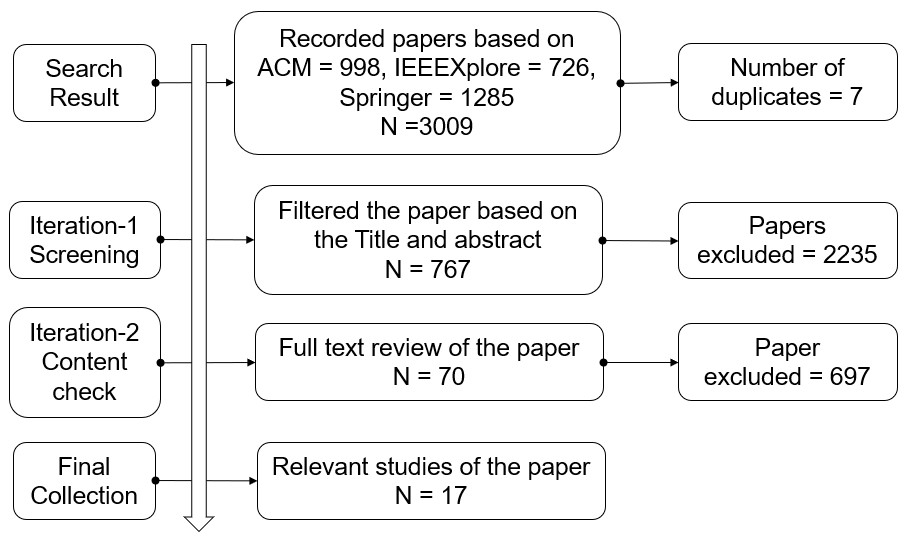}
    \caption{Illustration of filtering our Search Output}
    \label{fig:img1}
\end{figure}

We conducted our literature search in digital libraries like ACM, Springer, IEEEXplore, ScienceDirect, Wiley. We managed to extract relevant research papers from ACM, Springer, and IEEEXplore. We had to exclude ScienceDirect and Wiley from duplicates, and the results are relatively low compared to other digital libraries.  These two libraries have minor literature on both hardware and user interaction in context to VR Domain. We have considered only research articles only for our review study. By following our search strategy, we conducted an extensive search on the respective databases. As illustrated in Fig: \ref{fig:img1}, we extracted our search results in multiple levels by applying inclusion and exclusion criteria. As part of the initial search, we extracted 998 papers from ACM digital library, 726 papers from IEEEXplore, and 1285 papers from Springer Journal. We conducted a peer review and have reportedly removed seven duplicates across the search results. We extracted 3009 papers in total as part of the initial search.
As part of our first iteration of screening, we filtered papers based on title and abstract. We excluded around 2235 papers and considered 767 papers as part of this step. In the second iteration of screening, we conducted a full-text review of the paper based on the context of our search and have further filtered the search results to 70 papers by excluding 697 papers. We conducted a detailed study on the filtered papers regarding relevance, technique, and metrics as part of the final consideration. We managed to filter the results to 14 papers, and after further snowballing on related work \cite{experimentation}, 17 papers are finalized for our review study. All the authors have individually reproduced the search and applied the filters in respective iterations. All the authors have arrived at a similar conclusion towards the search results. All the supplementary material of our search iterations and review are made available for replicating our study \cite{supplement}.
\begin{table*}[ht]
\centering
\begin{tabular}{cl|lllccl}
\hline
\textbf{S. No} &
  \multicolumn{1}{c|}{\textbf{Application}} &
  \multicolumn{1}{c}{\textbf{\begin{tabular}[c]{@{}c@{}}Tracking \\ method\end{tabular}}} &
  \multicolumn{1}{c}{\textbf{Hardware setup}} &
  \multicolumn{1}{c}{\textbf{Working principle}} &
  \textbf{\begin{tabular}[c]{@{}c@{}}Locomotion\\  involved\end{tabular}} &
  \textbf{Year} &
  \multicolumn{1}{c}{\textbf{Ref.}} \\ \hline
\multirow{3}{*}{1} &
  \multirow{3}{*}{\begin{tabular}[c]{@{}l@{}}Sports \\ commodity\end{tabular}} &
  \begin{tabular}[c]{@{}l@{}}Color \\  tracking\end{tabular} &
  \begin{tabular}[c]{@{}l@{}}camera setup \\ \& HMD\end{tabular} &
  \begin{tabular}[c]{@{}l@{}}Image segmentation followed by\\  a classifier \& filtering algorithms\end{tabular} &
  \begin{tabular}[c]{@{}c@{}}rotation \& \\ translation\end{tabular} &
  2011 & \cite{1_wang_practical_2011}
   \\ \cline{3-8} 
 &
   &
  VR-STEP &
  \begin{tabular}[c]{@{}l@{}}IMU, pedometer\\  \& HMD\end{tabular} &
  \begin{tabular}[c]{@{}l@{}}Real-time pedometry to simulate \\ virtual locomotion\end{tabular} &
  \begin{tabular}[c]{@{}c@{}}rotation \& \\ translation\end{tabular} &
  2016 & \cite{2_tregillus_vr-step_2016}
   \\ \cline{3-8} 
 &
   &
  \begin{tabular}[c]{@{}l@{}}Walking by \\ Cycle\end{tabular} &
  \begin{tabular}[c]{@{}l@{}}VR strider, HMD \\ \& pressure sensors\end{tabular} &
  \begin{tabular}[c]{@{}l@{}}Strider action mapped to virtual \\ locomotion with pressure sensors\end{tabular} &
  \begin{tabular}[c]{@{}c@{}}rotational \& \\ axial\end{tabular} &
  2020 & 
\cite{3_freiwald_walking_2020}
   \\ \hline
\multirow{2}{*}{2} &
  \multirow{2}{*}{\begin{tabular}[c]{@{}l@{}}Exploratory \\ study\end{tabular}} &
  \begin{tabular}[c]{@{}l@{}}Omnidirectional \\ treadmill\end{tabular} &
  \begin{tabular}[c]{@{}l@{}}treadmill, camera\\  setup \& HMD\end{tabular} &
  \begin{tabular}[c]{@{}l@{}}Treadmill coupled with motor\\  with the camera setup\end{tabular} &
  \begin{tabular}[c]{@{}c@{}}rotation \& \\ translation\end{tabular} &
  2013 & \cite{4_camarinha-matos_omnidirectional_2013}
   \\ \cline{3-8} 
 &
   &
  Object tracking &
  \begin{tabular}[c]{@{}l@{}}camera setup\\  \& HMD\end{tabular} &
  \begin{tabular}[c]{@{}l@{}}3D human model rasterized\\  view using CUDA-OpenGL\end{tabular} &
  \begin{tabular}[c]{@{}c@{}}rotation \& \\ translation\end{tabular} &
  2014 & \cite{5_chmielewski_mixing_2014}
   \\ \hline
\multirow{2}{*}{3} &
  \multirow{2}{*}{\begin{tabular}[c]{@{}l@{}}Cognitive \\ study\end{tabular}} &
  \begin{tabular}[c]{@{}l@{}}Suspended \\ walking\end{tabular} &
  IMU \& HMD &
  \begin{tabular}[c]{@{}l@{}}3-axis IMU attached to the leg \\  as a step counter\end{tabular} &
  \begin{tabular}[c]{@{}c@{}}rotational \& \\ axial\end{tabular} &
  2013 & \cite{6_anacleto_suspended_2013}
   \\ \cline{3-8} 
 &
   &
  Walking in Place &
  \begin{tabular}[c]{@{}l@{}}OptiTrack system \\ \& HMD\end{tabular} &
  \begin{tabular}[c]{@{}l@{}}The height of the foot is mapped\\  to data in a virtual environment\end{tabular} &
  \begin{tabular}[c]{@{}c@{}}rotational \& \\ axial\end{tabular} &
  2013 & \cite{7_hutchison_new_2013}
   \\ \hline
\multirow{2}{*}{4} &
  \multirow{2}{*}{\begin{tabular}[c]{@{}l@{}}Multi-user\\ application\end{tabular}} &
  \begin{tabular}[c]{@{}l@{}}Infrared tracking \\ method\end{tabular} &
  \begin{tabular}[c]{@{}l@{}}IR camera setup \\ \& HMD\end{tabular} &
  \begin{tabular}[c]{@{}l@{}}Use of IR cameras to track markers\\  on user to calculate the position\end{tabular} &
  \begin{tabular}[c]{@{}c@{}}rotation \& \\ translation\end{tabular} &
  2017 & \cite{8_xu_multi-target_2017}
   \\ \cline{3-8} 
 &
   &
  \begin{tabular}[c]{@{}l@{}}Multi-user \\ tracking\end{tabular} &
  \begin{tabular}[c]{@{}l@{}}camera setup\\  \& HMD\end{tabular} &
  \begin{tabular}[c]{@{}l@{}}Depth images captured from Kinect \\ sensors and processed using a model\end{tabular} &
  \begin{tabular}[c]{@{}c@{}}rotation \& \\ translation\end{tabular} &
  2019 & \cite{9_bicho_markerless_2019}
   \\ \hline
\multirow{3}{*}{5} &
  \multirow{3}{*}{\begin{tabular}[c]{@{}l@{}}Redirected\\ motion\end{tabular}} &
  NaviChair &
  \begin{tabular}[c]{@{}l@{}}motion cued \\ chair \& HMD\end{tabular} &
  \begin{tabular}[c]{@{}l@{}}Leaning motion on the chair simulates \\ an action in the virtual environment\end{tabular} &
  \begin{tabular}[c]{@{}c@{}}rotational \& \\ axial\end{tabular} &
  2015 & \cite{10_kitson}
   \\ \cline{3-8} 
 &
   &
  Virtusphere &
  \begin{tabular}[c]{@{}l@{}}suspended sphere \\ setup \& HMD\end{tabular} &
  \begin{tabular}[c]{@{}l@{}}Freely suspended spherical frame\\  simulates action in a virtual space\end{tabular} &
  \begin{tabular}[c]{@{}c@{}}rotation \& \\ translation\end{tabular} &
  2015 & \cite{11_nabiyouni}
   \\ \cline{3-8} 
 &
   &
  \begin{tabular}[c]{@{}l@{}}Tapping in \\ Place\end{tabular} &
  \begin{tabular}[c]{@{}l@{}}walking pad, IMU\\ \& control system\end{tabular} &
  \begin{tabular}[c]{@{}l@{}}The walking pad acts as a navigation \\ key to simulate on IMU-based device\end{tabular} &
  \begin{tabular}[c]{@{}c@{}}rotational \& \\ axial\end{tabular} &
  2018 & \cite{12_hudak_walking_2018}
   \\ \hline
\multirow{3}{*}{6} &
  \multirow{3}{*}{Training} &
  \begin{tabular}[c]{@{}l@{}}Electromagnetic \\ tracking system\end{tabular} &
  \begin{tabular}[c]{@{}l@{}}EM tracking, Step \\ sensor \& HMD\end{tabular} &
  \begin{tabular}[c]{@{}l@{}}The user is localized based on the EM \\ track system using step sensor\end{tabular} &
  \begin{tabular}[c]{@{}c@{}}rotation \& \\ translation\end{tabular} &
  2016 & \cite{13_zank_improvements_2016}
   \\ \cline{3-8} 
 &
   &
  Elastic-Move &
  \begin{tabular}[c]{@{}l@{}}Suspended system,\\ Elastic belt \& HMD\end{tabular} &
  \begin{tabular}[c]{@{}l@{}}Uses elastic rope as force-based\\  feedback to simulate walking action\end{tabular} &
  \begin{tabular}[c]{@{}c@{}}rotational \& \\ axial\end{tabular} &
  2020 & \cite{14_yi_elastic-move_2020}
   \\ \cline{3-8} 
 &
   &
  \begin{tabular}[c]{@{}l@{}}Electromagnetic \\ tracking system\end{tabular} &
  \begin{tabular}[c]{@{}l@{}}Tx-HMD, cube-coil, \\ Rx-analyzer\end{tabular} &
  \begin{tabular}[c]{@{}l@{}}Calculates the location of central\\  magnetic coil using EM systems\end{tabular} &
  \begin{tabular}[c]{@{}c@{}}rotation \& \\ translation\end{tabular} &
  2020 & \cite{15_barai_outside-electromagnetic_2020}
   \\ \hline
\multirow{2}{*}{7} &
  \multirow{2}{*}{Simulation} &
  \begin{tabular}[c]{@{}l@{}}Acoustic position \\ tracking method\end{tabular} &
  \begin{tabular}[c]{@{}l@{}}acoustic Tx-Rx\\  setup \& HMD\end{tabular} &
  \begin{tabular}[c]{@{}l@{}}Uses acoustic sensing using pair of \\ speakers to track users location\end{tabular} &
  \begin{tabular}[c]{@{}c@{}}rotational \& \\ axial\end{tabular} &
  2018 & \cite{16_al_zayer_stereotrack_2018}
   \\ \cline{3-8} 
 &
   &
  \begin{tabular}[c]{@{}l@{}}Infrared tracking\\  method\end{tabular} &
  \begin{tabular}[c]{@{}l@{}}IR LED's setup \& \\ HMD\end{tabular} &
  \begin{tabular}[c]{@{}l@{}}The customized camera captures the\\  position of IR LEDs using a filter\end{tabular} &
  \begin{tabular}[c]{@{}c@{}}rotation \& \\ translation\end{tabular} &
  2020 & \cite{17_brooks_positional_2020}
   \\ \hline
\end{tabular}
\caption{Finalized studies for the review }
\label{tab:table2}
\end{table*}

\section{Discussion} \label{4}
By considering the finalized research papers, we conducted an elaborated study to record our findings. In this section, we discuss our insights in regards to respective research questions as follows: 

\begin{tcolorbox}[]
\textbf{RQ1}-\textit{What types of motion tracking methods are operated for head tracking by an HMD for VR applications?}
\end{tcolorbox}

Table \ref{tab:table2} illustrates the HMDs transformed to 6-DOF support with respective details on their hardware setup, working principle of the underlying motion tracking method to facilitate locomotion in VR applications. It also categorizes these HMDs based on the target VR application. Table \ref{tab:table2} also provides the year of publication along with its reference. We observe that the motion tracking methods are enhanced by hardware transformation largely for targeted applications like training, simulation, and multi-user application. All these hardware transformations are scaled and scoped to HMDs without any external haptic support. In almost all cases, the HMDs primarily supported locomotion techniques like rotational, translation, and axial. We further discuss the effectiveness of the underlying hardware of these transformed HMDs as part of RQ3. 
\\
\newline
\begin{tcolorbox}[]
\textbf{RQ2}: \textit{What are the hardware components required for transforming a VR HMD from 3-DOF to 6-DOF motion tracking?}
\end{tcolorbox}

Considering the insights from the reviewed papers, we address this research question by proposing a taxonomy of locomotion techniques based on hardware support. The underlying hardware is used to conduct locomotion through motion tracking methods in a given HMD. The motive behind illustrating the taxonomy is to help VR practitioners establish a relationship between the hardware needs of an HMD and choose a suitable locomotion method. Lisa Prinz et al. conducted an initial review of primary studies that involved different taxonomies related to locomotion in VR \cite{marie}. Previously proposed taxonomies are based on parameters like walking, redirection, teleportation, haptics, hand gestures, and materialistic feedback  \cite{t1} \cite{t2}\cite{t3}. These proposed taxonomies of locomotion are either human-centered or software-centered. They do not factor in the customization of HMDs. Data captured in our review study helped classify the locomotion techniques based on motion tracking by considering the customized HMD hardware.

\begin{figure*}[ht]
    \centering
    \includegraphics[scale=0.52]{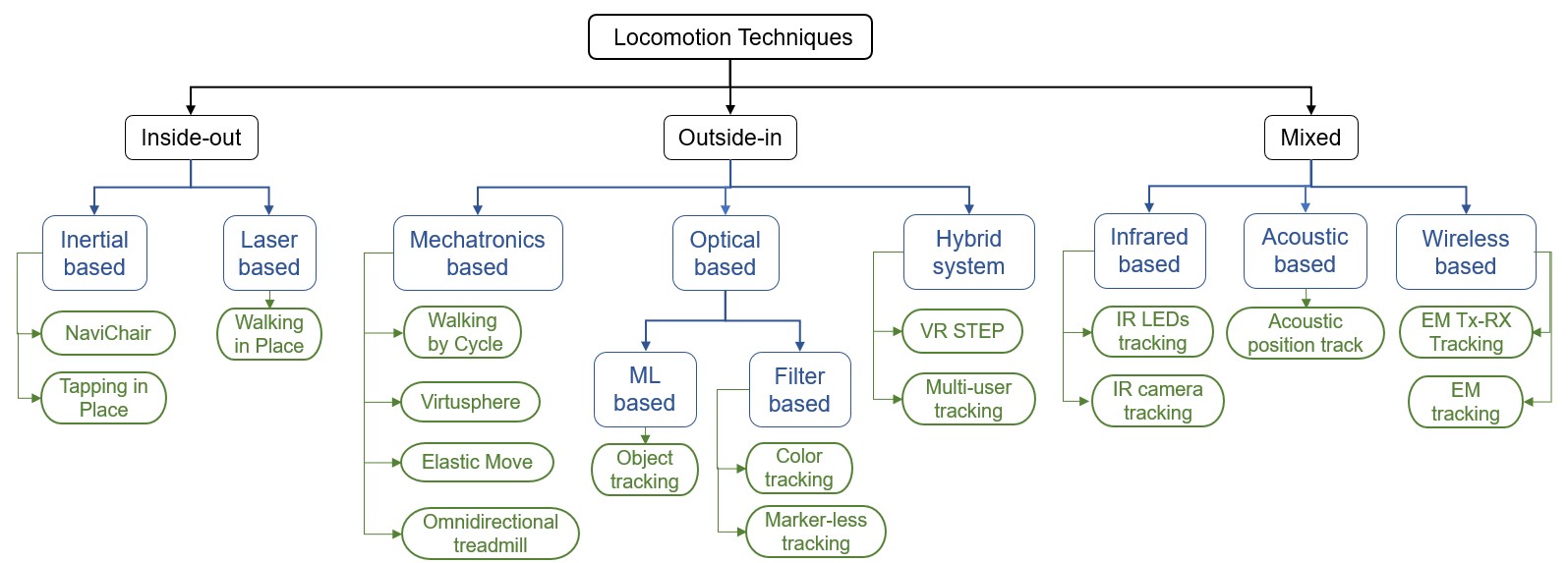}
    \caption{Taxonomy of the locomotion techniques based on motion tracking hardware}
    \label{fig:tax}
\end{figure*}

The characteristics of the hardware system can be defined by the performance of sensors \cite{t4}, actuators \cite{t5}, control system, and the processing unit \cite{nabiyouni_taxonomy_2016}. Considering these hardware characteristics, we propose a taxonomy for locomotion techniques as illustrated in Fig \ref{fig:tax}. Based on captured review information, we classify the hardware-based locomotion techniques for HMDs into three main categories as shown in figure. They are \textit{inside-out, outside-in,} and \textit{mixed tracking}. As part of our initial classification, we considered parameters like DOF, peripherals, and transfer function for taxonomy. However, we limited our taxonomy to tracking device's position only, as it will extensively help VR practitioners choose the best locomotion technique for their respective HMD. As shown in Fig \ref{fig:tax}, the boxes in blue are types of respective locomotion techniques categorized as Inside-out, Outside-In, and Mixed. The boxes in green are the instances or examples of these respective types listed in blue boxes. For example, \textit{'Navichair'} and \textit{'Tapping in Place'} are instances of Inertial based locomotion methods supported by the underlying hardware categorized as Inside-Out tracking.
\\
\newline
\textbf{Inside-Out tracking} - The sensors are located on the hardware (HMD) or the user peripherals as part of the Inside-Out tracking. Based on our review, three out of seventeen (17.6\%) locomotion techniques belong to inside-out tracking. The inside-out tracking can be further divided into two groups based on the hardware description: inertial and laser-based tracking. Inertial tracking involves analysis based on Inertial-Measurement-Unit (IMU) to track the rotational and translational locomotion in VR HMDs \cite{12_hudak_walking_2018}. They are primarily mounted to track the head movements of the user. However, they are not immune to external noise and need a compensation mechanism to counter motion sickness-related problems \cite{10_kitson}. On the other hand, laser-based systems like lighthouse tracking use rectangular base stations as reference points to accurately track the user's position, and orientation \cite{7_hutchison_new_2013}. These are user-centric, and the data points are gathered based on user stimuli. 
\\
\newline
\textbf{Outside-In tracking} - The sensors are placed externally, preferably in a stationary position, and are not administered directly on the user's device as part of Outside-In tracking. This locomotion technique is observed to be dominant in practice with nine out of seventeen (52.9\%). As per our review, the outside-in tracking can be further classified into three groups. They are mechatronics, optical and hybrid systems. The studies involving redirected actions using mechanical instruments like treadmill, cycle, hamster ball, or suspended walking using elastic can be categorized as mechatronics systems \cite{4_camarinha-matos_omnidirectional_2013} \cite{3_freiwald_walking_2020} \cite{11_nabiyouni}. Locomotion that involves camera-based tracking are categorized into optical systems. Considering the working principle of the camera setup, the optical systems can be further subdivided into two groups: filter and machine-learning (ML) based techniques. The filter-based techniques used color tracking or projection-based detection methods \cite{1_wang_practical_2011}, while object tracking and data prediction are achieved using neural networks \cite{7_hutchison_new_2013}. Further studies on simulation of motion using mechanical instruments followed by prediction model using ML tools are classified as hybrid systems \cite{5_chmielewski_mixing_2014}\cite{9_bicho_markerless_2019}.
\\
\newline
\textbf{Mixed tracking} - Further studies have found to be following mixed-methods i.e., they employ both Inside-Out and Outside-In tracking. We categorized them as Mixed Tracking methods. Five out of seventeen (29.4\%) locomotion techniques use the user as a receiver and an external point as a transmitter to track position and orientation. Considering the working principle of these systems, we further divided them into three groups: infrared, acoustic, and wireless-tracking systems. Infrared tracking involves locomotion using remote communication utilizing IR cameras \cite{8_xu_multi-target_2017}. In some cases, position tracking was conducted using the IR LED's on the user body using an external camera\cite{17_brooks_positional_2020}. The tracking involving Electro-Magnetic(EM) transmission using base station \cite{13_zank_improvements_2016}\cite{15_barai_outside-electromagnetic_2020}. These are classified as wireless tracking methods. A novel tracking method for localization using audio inputs(stereo speakers) is categorized as acoustic tracking \cite{16_al_zayer_stereotrack_2018}.

\begin{tcolorbox}[]
\textbf{RQ3}: \textit{What are the metrics practiced to evaluate the effectiveness of motion tracking methods after transforming the VR HMD from 3-DOF to 6-DOF?}
\end{tcolorbox}

Table \ref{tab:tab3} illustrates the evaluation details of the effectiveness of motion tracking methods after transforming the VR HMD from 3-DOF to 6-DOF. Table \ref{tab:tab3} is a matrix table with the measures listed in the first row and first column grouped by the citation of the respective paper that employs these measures. The measure listed in the first row is the empirical method used to conduct the evaluation. The measure listed in the first column describes the metrics gathered to understand the effectiveness of the respective motion tracking method after transforming the HMD from 3-DOF to 6-DOF. We address this research question we categorized the research contributions based on type of the empirical study. We also present the underlying participant experiences and metrics used by VR practitioners. 
\\
\newline
\textbf{Exploratory Studies -} The following evaluations are conducted using Sports VR applications like basketball, Jogging In-place, and Walking-by-Cycling. 
\\
\newline
Robert Wang et al. used a two-camera system in real-world indoor and outdoor environments for various activities and lighting conditions. They conducted a focused group evaluation, including basketball players with 3-DOF HMD wearing colored t-shirts set to be detected by these two camera systems. This camera input is sent as locomotion feedback to the HMD.  They captured footage in a dimly lit indoor basketball court through a glass panel of a squash court. Their study was easy to set up with less use of additional lights or equipment \cite{1_wang_practical_2011}. They managed to capture Position accuracy, Drift measure, Precision, and System stability to understand the effectiveness of this setup. 

Sam Tregillus et al. have conducted a comparative study of all Walk-In-Place (WIP) methods as part of their study. Several such studies exist that compared WIP with joystick-based virtual locomotion. However, Sam Tregillus et al. presumed that comparisons of WIP methods involving extensive instrumentation are not helpful as the users of mobile VR do not access such instrumentation. They created VR-STEP that is hands-free and requires no instrumentation. It is more meaningful to compare its performance with another hands-free navigation method like \textit{``look down to move''} (LDTM), widely used in several VR apps. Here, the users toggle a button at their feet by briefly looking down at it, then back it up. When activated, the user will move with a fixed horizontal velocity in the direction of their gaze. Similar to other WIP evaluations, they compared VR-STEP to LDTM by having users perform several navigation tasks \cite{2_tregillus_vr-step_2016}. They used a Jogging-Inplace VR sport application to conduct the study. They captured track resolution, Motion-to-Photon (M2P) latency and conducted a Statistical analysis to evaluate their 3-DOF HMD. 

Jann Freiwald et al. conducted a focused group study using their 3-DOF HMD with a non-swiveling chair setup. 20 participants (Mean = 30.6, SD = 6.82, 7 female) took part in the experiment. The mean time per participant was about 60 minutes. They build a Walking-by-Cycling sports application using Unity3D. The rendered scene is run on HTC Vive Pro and a non-swiveling chair to seat the participants within the tracking space for the bike and joystick conditions. The participants are asked to stand for the teleportation condition. Standing was required to let the participants use their full head and body proprioception for angular estimation as a baseline to test against. Depending on the condition, they used an Xbox One Controller or the HTC Vive Wand for input \cite{3_freiwald_walking_2020}. Response Time, Force Feedback, Track resolution, and Drift measure to evaluate the effectiveness of the locomotion. 
\\
\newline
\textbf{Computational Studies -} The following evaluations are Computational and Parametric studies using omni-directional treadmill and object tracking.
\\
\newline
Razvan Boboc et al. conducted an exploratory study to examine locomotion using the neural-network-based algorithm. Six participants took part in this study in a virtual environment over an Omni-directional treadmill simulating the sense of steep in the hill or slove in a cave using a 3-DOF HMD. The position and orientation of the user's feet are captured using the motion tracking system. These are input features of the algorithm. Later, the data is processed to extract the angle between the foot and tibia; the foot's orientation regards the reference position for the right and left rotation of the Omni-directional treadmill. Using Matlab Simulink, the authors have modeled positions of the motors related to a reference factor and a parameter for the scenario, i.e., the model is used to tilt the platform depending on the inclination angle of hills scenarios. For choosing the number of neurons, the neural network is trained in Matlab. The number of neurons in the hidden layer is reduced until the error can be accepted. \cite{4_camarinha-matos_omnidirectional_2013}. They captured metrics like position and rotation accuracy, precision. They also conducted statistical analysis to understand the significance of their data points. 

Boguslaw Rymut et al. conducted an exploratory computational study on object tracking in VR Sence based on a real-time multiuser interface using a 3-DOF HMD. Their algorithm's performance has been evaluated on sequences with walking persons. They demonstrated that the average speed-up of GPU over CPU is about 7.5. The overall time taken by PSO searching for the best matching image is far shorter than the time needed for evaluation of the fitness function, which is about 0.9 ms.\cite{5_chmielewski_mixing_2014}. They captured track resolution, portability, precision, motion-to-photon (M2P) latency, and response time metrics. They conducted statistical analysis to understand the significance of their data points. 
\\
\newline
\textbf{Cognitive Studies -} The following evaluations are Cognitive studies using Suspended walking and Walking-in-Place.
\\
\newline
Benjamin Franks et al. conducted a focused group assessment with 18 test persons aged 22 to 30 years (average 26.3). Seven of the participants were male, eleven of them were female. The participants are instructed to play a customized level in the game 'Portal 2' using a 3-DOF HMD. The level consisted of an obstacle course specifically designed for the experiment using a level editor  \cite{6_anacleto_suspended_2013}. After the game completion, they captured the feedback to evaluate the hardware based on drift measure, track resolution, force feedback, and system stability. The response revealed that none of the participants found the suspension setup most comfortable.  Some participants criticized that contrary to WIP, the setup restricted the backward locomotion.

Luis Bruno et al. conducted a cognitive study using the OptiTrack motion system. The experiment involved was divided into three segments. Prior to the test,  the participants are asked to take a pre-test questionnaire to gather each participant's demographic and navigation skills data. The participant was made comfortable with the environment by performing travel and stopping tasks. The actual test involved traveling all nine paths and making stops before each target as early as possible \cite{7_hutchison_new_2013}. The path was repeated in case of system error or difficulty. After the task, the participants were subjected to the post-test questionnaire to get feedback about their experience. The metrics evaluated using the experiment are drift measure, switching rate, interface network, and response time.
\\
\newline
\textbf{Multi-User Applications -} The following evaluations are Multi-user applications using infrared markers and markerless multi-view tracking.
\\
\newline
Wenhui Xu et al. performed a focused group assessment using two individual experiments based on the parameters \cite{8_xu_multi-target_2017}. The first experiment uses three infrared cameras with a resolution of 1280 x 720 pixels. The participant with the LED module is made to stand three meters away from the cameras. The test was conducted for varying positions and orientations of the participant. The observations suggested that fluctuations do not alter the behavior of the VR display. The second experiment tests the data refresh rate of the system. They experimented with varying resolutions of display for three cameras and a four-camera setup. The results show that resolutions have a slight effect on the accuracy in the indoor environment. The metrics captured in both experiments are positional accuracy, rotational accuracy, precision, response time, and system stability.

Dylan Bicho et al. conducted a user-centered study and proposed a method using four Microsoft Kinect sensors \cite{9_bicho_markerless_2019}. The intent was to capture the locomotion of the participant in the different orientations using a 3-DOF HMD. The participant was asked to perform individual tasks like walking straight, following a square-shaped path in a closed-loop, moving in a random path with sudden body or head motions or standing stationary on a single leg with arms wide open. The metrics captured in the experiment are drift measure, track resolution, portability or customization, learning curve, and switching rate.
\\
\newline
\textbf{Redirected Motion -} The following evaluations are conducted using redirected motion applications like NaviChair, Vitrusphere and Tapping-in-Place. 
\\
\newline
Alexandra Kitson et al. conducted a user-centred study to evaluate the factors responsible for motion sickness in NaviChair \cite{10_kitson}. The participants tested for two locomotion techniques in the experiment. The first experiment involved motion using a user-powered swivel chair called NaviChair. The participants can move forward by tilting the chair forward and rotating the chair to rotate in the virtual environment. The second experiment involved a similar set of locomotions, but the joystick was used as the input device. The observations concluded that NaviChair did not help the participants localize and adjust to the virtual environment. The metrics evaluated using both experiments are drift measure, learning curve, interface network, M2P latency and statistical analysis.

Mahdi Nabiyouni et al. conducted a simulation-based study using a suspended sphere called Virtusphere \cite{11_nabiyouni}. The user's walking is mapped to a viewpoint translation in the virtual scene. They performed a comparative analysis of semi-natural techniques like Virtusphere with an entirely natural technique like walking and artificial technique using a game controller. The analysis suggested that the Virtusphere method was significantly slower and less accurate than the other two techniques. The parameters involved for the comparative analysis were drift measure, rotation accuracy, force feedback and system stability.

Marian Hudak et al. did the comparative analysis using a customized CAVE setup \cite{12_hudak_walking_2018}. They used the 250-degree panoramic view to simulate the surface of the cave as a virtual environment. The participant was asked to perform standard walking and rotation movements. The forward movement was represented on the central tile, and step-aside rotation tiles represented pan rotation movements. The metrics captured in the analysis were position accuracy, precision, noise immunity, interface network and response time.
\begin{table*}[ht]
\centering
\begin{tabular}{|c|c|c|c|c|c|c|c|}
\hline
\diagbox{\textbf{Metrics}}{\textbf{Study}} &
  \multicolumn{1}{c|}{\begin{tabular}[c]{@{}c@{}}\textbf{Focused}\\\textbf{Group studies}\end{tabular}} &
  \multicolumn{1}{c|}{\begin{tabular}[c]{@{}c@{}}\textbf{Cognitive}\\\textbf{Studies}\end{tabular}} &
  \multicolumn{1}{c|}{\begin{tabular}[c]{@{}c@{}}\textbf{Comparative}\\ \textbf{Studies}\end{tabular}} &
  \multicolumn{1}{c|}{\begin{tabular}[c]{@{}c@{}}\textbf{Compute}\\ \textbf{Studies}\end{tabular}} &
  \multicolumn{1}{c|}{\begin{tabular}[c]{@{}c@{}}\textbf{User-centered}\\ \textbf{Studies}\end{tabular}} &
  \multicolumn{1}{c|}{\begin{tabular}[c]{@{}c@{}}\textbf{Simulation}\\ \textbf{Studies}\end{tabular}} &
  \multicolumn{1}{c|}{\begin{tabular}[c]{@{}c@{}}\textbf{Experimental}\\ \textbf{Studies}\end{tabular}} \\ \hline
\begin{tabular}[c]{@{}c@{}}\textbf{Position}\\ \textbf{Accuracy}\end{tabular}  &\cite{1_wang_practical_2011}\cite{8_xu_multi-target_2017}\cite{16_al_zayer_stereotrack_2018}  &\cite{7_hutchison_new_2013}  &\cite{12_hudak_walking_2018}\cite{17_brooks_positional_2020}  &\cite{4_camarinha-matos_omnidirectional_2013}\cite{15_barai_outside-electromagnetic_2020}  & \cite{8_xu_multi-target_2017} &\cite{12_hudak_walking_2018}  &\cite{7_hutchison_new_2013}\cite{8_xu_multi-target_2017} \cite{13_zank_improvements_2016} \\ \hline
\begin{tabular}[c]{@{}c@{}}\textbf{Drift} \\ \textbf{measure}\end{tabular}       &\cite{1_wang_practical_2011}\cite{3_freiwald_walking_2020}\cite{6_anacleto_suspended_2013}\cite{7_hutchison_new_2013} & \cite{6_anacleto_suspended_2013}\cite{7_hutchison_new_2013}  & \cite{13_zank_improvements_2016}\cite{17_brooks_positional_2020} & \cite{15_barai_outside-electromagnetic_2020}&\cite{8_xu_multi-target_2017}\cite{9_bicho_markerless_2019}\cite{10_kitson} &\cite{11_nabiyouni}  &\cite{8_xu_multi-target_2017}\cite{10_kitson}\cite{13_zank_improvements_2016} \\ \hline

\begin{tabular}[c]{@{}c@{}}\textbf{Rotation}\\ \textbf{Accuracy}\end{tabular}  &\cite{8_xu_multi-target_2017}\cite{14_yi_elastic-move_2020}\cite{16_al_zayer_stereotrack_2018}\cite{17_brooks_positional_2020}  &  &\cite{13_zank_improvements_2016}\cite{17_brooks_positional_2020}  &\cite{4_camarinha-matos_omnidirectional_2013}\cite{14_yi_elastic-move_2020}    & \cite{8_xu_multi-target_2017}\cite{10_kitson}\cite{11_nabiyouni} &\cite{10_kitson}\cite{11_nabiyouni}  &\cite{8_xu_multi-target_2017}\cite{10_kitson}\cite{13_zank_improvements_2016} \\ \hline

\begin{tabular}[c]{@{}c@{}}\textbf{Track}\\ \textbf{Resolution}\end{tabular}  &\cite{3_freiwald_walking_2020}\cite{6_anacleto_suspended_2013}\cite{16_al_zayer_stereotrack_2018}  &\cite{6_anacleto_suspended_2013}\cite{9_bicho_markerless_2019}  &\cite{2_tregillus_vr-step_2016}\cite{12_hudak_walking_2018}\cite{15_barai_outside-electromagnetic_2020}  &\cite{5_chmielewski_mixing_2014}  &\cite{9_bicho_markerless_2019}    &  &\cite{9_bicho_markerless_2019}  \\ \hline

\begin{tabular}[c]{@{}c@{}}\textbf{Portability}\end{tabular}                                    &  &\cite{9_bicho_markerless_2019}  &\cite{5_chmielewski_mixing_2014}\cite{17_brooks_positional_2020}  &\cite{5_chmielewski_mixing_2014}    &\cite{9_bicho_markerless_2019}  &  &\cite{9_bicho_markerless_2019}  \\ \hline

\begin{tabular}[c]{@{}c@{}}\textbf{Learning} \\ \textbf{curve}\end{tabular}      & \cite{14_yi_elastic-move_2020} &\cite{9_bicho_markerless_2019}  &  &\cite{14_yi_elastic-move_2020}   &\cite{9_bicho_markerless_2019}\cite{10_kitson}  &\cite{10_kitson}  &\cite{9_bicho_markerless_2019}\cite{10_kitson}\cite{14_yi_elastic-move_2020}  \\ \hline

\begin{tabular}[c]{@{}c@{}}\textbf{Precision}\end{tabular}                                                    &\cite{1_wang_practical_2011}\cite{8_xu_multi-target_2017}  &  &\cite{5_chmielewski_mixing_2014}\cite{12_hudak_walking_2018}  &\cite{4_camarinha-matos_omnidirectional_2013}\cite{5_chmielewski_mixing_2014}    &\cite{8_xu_multi-target_2017}  &\cite{12_hudak_walking_2018}  &\cite{4_camarinha-matos_omnidirectional_2013}\cite{8_xu_multi-target_2017}  \\ \hline

\begin{tabular}[c]{@{}c@{}}\textbf{Switching} \\ \textbf{Rate}\end{tabular}      & \cite{7_hutchison_new_2013}\cite{8_xu_multi-target_2017}\cite{17_brooks_positional_2020} & \cite{7_hutchison_new_2013}\cite{9_bicho_markerless_2019} &\cite{17_brooks_positional_2020}  &    &\cite{8_xu_multi-target_2017}\cite{9_bicho_markerless_2019}  &  &\cite{7_hutchison_new_2013}\cite{8_xu_multi-target_2017}\cite{9_bicho_markerless_2019}  \\ \hline

\begin{tabular}[c]{@{}c@{}}\textbf{Force} \\ \textbf{feedback}\end{tabular}      &\cite{3_freiwald_walking_2020}\cite{6_anacleto_suspended_2013}\cite{14_yi_elastic-move_2020} &\cite{6_anacleto_suspended_2013} &  &\cite{14_yi_elastic-move_2020}    &\cite{11_nabiyouni}  &\cite{11_nabiyouni}  &\cite{14_yi_elastic-move_2020}  \\ \hline

\begin{tabular}[c]{@{}c@{}}\textbf{Noise} \\ \textbf{Immunity}\end{tabular}      & \cite{13_zank_improvements_2016}\cite{14_yi_elastic-move_2020}\cite{16_al_zayer_stereotrack_2018} &  &\cite{12_hudak_walking_2018}\cite{13_zank_improvements_2016} &\cite{14_yi_elastic-move_2020}\cite{15_barai_outside-electromagnetic_2020}    &  &\cite{12_hudak_walking_2018}  &\cite{13_zank_improvements_2016}\cite{14_yi_elastic-move_2020}\cite{15_barai_outside-electromagnetic_2020}  \\ \hline

\begin{tabular}[c]{@{}c@{}}\textbf{Interface}\\ \textbf{network}\end{tabular}                                                    &\cite{7_hutchison_new_2013}\cite{12_hudak_walking_2018}\cite{16_al_zayer_stereotrack_2018}\cite{17_brooks_positional_2020}  &\cite{7_hutchison_new_2013}  &\cite{5_chmielewski_mixing_2014}\cite{12_hudak_walking_2018}\cite{17_brooks_positional_2020}
& \cite{5_chmielewski_mixing_2014}\cite{14_yi_elastic-move_2020}   & \cite{10_kitson} &\cite{10_kitson}\cite{12_hudak_walking_2018}  &\cite{7_hutchison_new_2013}\cite{10_kitson}\cite{14_yi_elastic-move_2020}  \\ \hline

\begin{tabular}[c]{@{}c@{}}\textbf{Power}\\ \textbf{Analysis}\end{tabular}       &\cite{16_al_zayer_stereotrack_2018}  &  &\cite{13_zank_improvements_2016}\cite{15_barai_outside-electromagnetic_2020}  &\cite{15_barai_outside-electromagnetic_2020}    &  &  &\cite{13_zank_improvements_2016}\cite{15_barai_outside-electromagnetic_2020}  \\ \hline

\begin{tabular}[c]{@{}c@{}}\textbf{Response}\\ \textbf{Time}\end{tabular}        &\cite{3_freiwald_walking_2020}\cite{7_hutchison_new_2013}\cite{8_xu_multi-target_2017}  & \cite{7_hutchison_new_2013} &\cite{5_chmielewski_mixing_2014}\cite{12_hudak_walking_2018}  & \cite{5_chmielewski_mixing_2014}   &\cite{8_xu_multi-target_2017}  & \cite{12_hudak_walking_2018} & \cite{7_hutchison_new_2013}\cite{8_xu_multi-target_2017} \\ \hline

\begin{tabular}[c]{@{}c@{}}\textbf{System}\\ \textbf{Stability}\end{tabular}     &\cite{1_wang_practical_2011}\cite{8_xu_multi-target_2017}\cite{17_brooks_positional_2020}  &  & \cite{13_zank_improvements_2016}\cite{17_brooks_positional_2020} &    &\cite{8_xu_multi-target_2017}\cite{11_nabiyouni}  &\cite{11_nabiyouni}  & \cite{8_xu_multi-target_2017}\cite{13_zank_improvements_2016} \\ \hline

\begin{tabular}[c]{@{}c@{}}\textbf{Statistical}\\ \textbf{Analysis}\end{tabular} &\cite{16_al_zayer_stereotrack_2018}  &  &\cite{2_tregillus_vr-step_2016}\cite{5_chmielewski_mixing_2014}\cite{13_zank_improvements_2016} &\cite{4_camarinha-matos_omnidirectional_2013}\cite{5_chmielewski_mixing_2014}\cite{15_barai_outside-electromagnetic_2020}    &\cite{10_kitson}  &\cite{10_kitson}  &\cite{10_kitson}\cite{13_zank_improvements_2016}\cite{15_barai_outside-electromagnetic_2020}  \\ \hline
\begin{tabular}[c]{@{}c@{}}\textbf{M2P}\\\textbf{latency}\end{tabular}  &  &  &\cite{2_tregillus_vr-step_2016}\cite{5_chmielewski_mixing_2014}  &\cite{5_chmielewski_mixing_2014}    &\cite{10_kitson}  & \cite{10_kitson}  & \cite{10_kitson}  \\ \hline
\end{tabular}
\caption{Evaluation and metrics for the locomotion techniques}
\label{tab:tab3}
\end{table*}
\\
\newline
\textbf{Training -} The following evaluations are conducted using training applications like Electromagnetic tracking and Elastic-move. 
\\
\newline
Markus Zank et al. did a comparative study for walking in a straight line with the old autonomous tracking system \cite{13_zank_improvements_2016}. The comparison was made primarily on parameters such as signal for foot movements, base's movement along the walking direction and movements in upward and sideward direction. The metrics evaluated in the study were drift measure, rotational accuracy, noise and power analysis, system stability and statistical analysis.

Da-Chung Yi et al. conducted a focused group study to evaluate the elastic-move system \cite{14_yi_elastic-move_2020}. The experiment was performed using Simulation Sickness Questions (SSQs) to validate the use of Elastic-Rope and Elastic-Box in VR. The participant was asked to move from a base point to the route ends in a customized virtual environment. The duration of the entire experiment is about 15 minutes. The metrics evaluated for successful motion tracking were rotation accuracy, learning curve, force feedback and noise immunity.

Shantanu Barai et al. did an experimental analysis to localize the participant using an EM-based Tx-Rx system \cite{15_barai_outside-electromagnetic_2020}. The EM transmitter was set up at a stationary point, and the secondary coil was mounted on the custom 3-DOF HMD. The experiment is carried out with the varied location of the secondary coil in the X-Y plane. The Z-value is kept constant and is compensated by transformation. The metrics captured using the experiment are position accuracy, drift measure, noise immunity, power and statistical analysis.
\\
\newline
\textbf{Simulation Studies -} The following evaluations are conducted using Simulation-based VR applications like tracking using acoustic and infrared LED setup. 
\\
\newline
Majed Al Zayer et al. used two acoustic speakers as the communication model to track the user's location who wears a 3-DOF HMD. They conducted a focused group evaluation for StereoTrack with a smartphone-based microphone to record ultrasonic tones. Multiple tones with varied frequencies were played on three different speaker interfaces to analyze the values \cite{16_al_zayer_stereotrack_2018}. They succeeded in verifying the hardware by measuring positional accuracy, rotational accuracy for 180 degrees, noise immunity, power, and statistical analysis.

Rasmus Eklund et al. conducted a comparative study for infrared tracking systems based on tests inspired by the Brimijoin experiment. The experiment confirmed that the participants with 3-DOF HMDs slightly moved their heads back and forth at about 15 degrees compared to no head movement. They repeated the experiment to see if participants could notice the degree of externalization once they stopped moving their heads \cite{17_brooks_positional_2020}. They captured rotational accuracy, switching rate, interface network, and system stability to evaluate the performance of the hardware.
\\
\newline
\textbf{Strategizing Hardware Transformation:} We presume that the observations captured from the above research questions will provide VR practitioners a reasonable choice for hardware transformation of their 3-DOF HMD. We systematically structured our observations so that the VR practitioners can understand the prevailing practices and channelize their HMD needs by picking up the desired evaluation method and metric to judge their transformed HMD. We have ensured that our complied taxonomy is compact and easy to comprehend the locomotion techniques for better hardware transformation. 

\section{Threats to Validity} \label{5}
We discuss the following threats to the validity of our review study- \begin{itemize}[noitemsep,nolistsep]
\item \textbf{Conclusion Validity} - We conducted an extensive search and filtration of research papers to deduce our observations. Our observations are factual and are recorded in detail based on the published research articles. Our conclusions are genuine and can be replicated by repeating the study using our search protocol. 

\item \textbf{Internal Validity} - We worked with peer-reviewers who are experts in the VR domain to help us with search string finalization, filtration, review, and analysis. Throughout the review, we received constant feedback on our search strategy. The possibilities of a few primary studies being overlooked are limited. Of course, authors could make minor mistakes regarding the judgment of a research paper during the filtration process. The peer-researchers agree on search Strings. 

\item \textbf{Construct Validity} - Our review observations are captured directly from the finalized research papers. All the claims in the studied research papers are to be subjected to the respective papers' primary authors. Our review observations are actual and are firmly illustrated for our understanding. 

\item \textbf{External Validity} - We have made every attempt to conduct this review under a systematic review protocol. Results may differ if the search strategy and data extraction are renewed with a different protocol. However, we guarantee that one can replicate and generate our observations by following our search strategy. 
\end{itemize}

\section{Related Work} \label{6}
In the past couple of decades, research on locomotion in the virtual environment started gaining its pace. Initially, the navigation in the virtual world was restricted to 3-DOF. Eventually, researchers focused on the user's other locomotive or navigation feedback to enhance involvement and immersive experience in VR. Many motion tracking techniques were implemented based on actions like walking, steering, selection and manipulation. Al Zayer et al. surveyed multiple tracking methods and discussed their strengths, weaknesses and application to provide an overview for the researchers to apply a particular technique \cite{al_zayer_virtual_2020}. When it comes to movements that can be employed in VR using locomotion, there is a need for proper classification based on the body organs involved, the extent of the action and its repeatability. Mahdi Nabiyouni et al. proposed a taxonomy of walking based locomotion techniques in VR \cite{nabiyouni_taxonomy_2016}. This work with comparative analysis provides insight to the researchers and system designers into choosing walking techniques and performing experiments to evaluate them. Lisa Prinz et al. carried out a review and analysis of 29 papers providing locomotion techniques taxonomies. The work inspires the researchers to develop taxonomies in coming up with a novel tracking methodology \cite{marie}. 
\par Heni Cherni et al. conducted a review for 22 motion tracking methods from 2012 to 2019 and provided guidelines to choose the method based on the user's application. The research was based on the HCI aspect of the VR locomotion and proposed a taxonomy based on user body-centred, external peripheral and mixed methods. The role of user body-centred motions and their relation to motion sickness were some of the paper's significant contributions. VR locomotion and sickness induced by it needs to be evaluated and corrected. Thomas Gemert et al. highlighted some key components to quantify VR sickness and metrics to counter them \cite{van_gemert_evaluating_2021}. However, these reviews on VR locomotion do not discuss the device parameters and the hardware requirements for designing a particular motion tracking system. The data relating to the device's operating range, the ability of the sensors, and its relation to the target application are essential viewpoints for a researcher developing a novel tracking method. As a part of the review, we tried to address the hardware aspect for the heuristic replication of the method.
\section{Conclusion} \label{7}
Locomotion in a VR Scene plays a critical role in user engagement. Motion tracking methods play a crucial role in locomotion. The primary goal of this paper is to identify the motion tracking methods operated in HMDs for conducting effective locomotion. This led us to conduct a systematic literature review to analyze the trend adopted by VR practitioners on transforming their 3-DOF based HMDs into 6-DOF for improved motion tracking scoped to HMDs only. Our review study revealed that different motion tracking methods like color tracking, object tracking, walking in-place, tapping in-place, elastic-move, etc., are achieved directly using an HMD without any influence of external haptic controllers. Adopting identified hardware components in transforming the 3-DOF HMDs closer to 6-DOF may help VR practitioners run rich VR content with effective locomotion. We also constructed a taxonomy for VR practitioners to understand the hardware ecosystem of HMD based on locomotion techniques. This will help VR practitioners expand their current HMDs and upscale them to support new motion tracking methods for effective locomotion. We also recorded potential evaluation methods and their related metrics practiced while assessing the hardware transformation of HMDs in context to their performance and scale. Overall, we expect that our review study will help HMD developers to consider all available observations to build novel HMDs using distinct motion tracking methods for focused VR applications. 
\\
\newline
\textbf{Future Work: } This review helped us channelize the strengths of prevailing hardware transformation setups, evaluation practices, and potential metrics practiced by VR practitioners. We plan to use this review study to develop a customized motion tracking method for a novel 3-DOF HMD to develop a focused healthcare application. These healthcare VR applications are planned to be low-cost and have better ease of use. 
\acknowledgments{
The authors thank our peer-researchers and VR practitioners who took part of search quality assessment and helped executing search strategy through out this review study.
}
\bibliographystyle{ieeetr}
\bibliography{main}
\end{document}